\documentstyle [fleqn] {article}
\begin{document}

   \begin{titlepage}
 \begin{flushright} ZU--TH3/94   \end{flushright}
 \vspace*{2cm}
 \begin{center}
   {\huge LIGHT DEFLECTION IN PERTURBED  FRIEDMANN UNIVERSES}
    \vspace{0.2cm}  \\     {\bf \large Ruth Durrer}
    \vspace{1cm} {\large
   \\ Universit\"at Z\"urich, Institut f\"ur Theoretische Physik,
      Winterthurerstrasse ~190, \\ CH-8057 Z\"urich, Switzerland}
	\vspace{2cm}
 \end{center}

 \begin{abstract}
 In this letter a new formula for light deflection is derived using only
 physically observable concepts. The general result is specialized
 to cosmological
 perturbation theory and expressed in terms of gauge--invariant perturbation
 variables. The resulting scalar, vector and tensor equations are
 supplemented by simple examples for illustration. The gravity wave
 example may be of more than academic interest and even represent a new way
 to detect gravitational waves.
 \vspace{1cm} \end{abstract}

 \end{titlepage}

The propagation of photons in a perturbed Friedmann universe leads to
 perturbations in the cosmic microwave background  which have been
detected recently, and to perturbations of light from distant sources
like quasars. Several formulas to calculate the perturbation of
photon energies (Sachs--Wolfe effect) have been derived so far, but
 the deflection of light in the framework of cosmological perturbation
theory has not been generally investigated.

On large scales the geometry of our Universe can be approximated to a
precision of about $10^{-5}$ by a homogeneous and isotropic Friedmann
Lema\^{\i}tre spacetime. Deviations from homogeneity and isotropy may
thus be treated in first order perturbation theory.
Cosmological perturbation theory was first investigated by Lifshitz
\cite{Li}. Since only the perturbed geometry is physically observable, the
choice of a background universe on which perturbations are defined is
somewhat arbitrary and is called a choice of gauge. Only quantities
whose background contributions vanish are gauge independent \cite{d93}.
Physically measurable quantities can always be expressed in terms of such
gauge--invariant variables, which usually have a simple geometrical
meaning \cite{BDE}.

In this letter we want to derive a formula for light deflection in terms of
gauge--invariant variables.
Throughout, we choose the metric signature  $(-,+,+,+)$.

On their way from the last scattering surface into our antennas,
 microwave photons travel through a perturbed Friedmann geometry.
Thus, even if the photon temperature was completely  uniform at the
last scattering surface, we  receive it slightly perturbed  \cite{SW}.
In addition, photons
traveling through a perturbed universe are deflected. We now
calculate this deflection in first order perturbation theory.

Metrics  which are conformally equivalent,
$ d\tilde{s}^2 = a^2ds^2$,
have the same lightlike geodesics, only the corresponding affine parameters
are different. Since Friedmann---Lema\^{\i}tre models are conformally flat,
we may thus discuss the propagation of light in a
 perturbed Minkowski geometry.  This simplifies things greatly. We denote
the affine parameter by $\lambda$ and the tangent vector to the  geodesic by
$ n = dx/d\lambda$, $n^2  = 0$, with unperturbed values $n^0 =1$ and
$\mbox{\boldmath $n$}^2 =1$. The tangent vector of the
perturbed geodesic is given by
$(1,\mbox{\boldmath $n$})  +\delta n$. The geodesic equation for the metric
\[ ds^2 = (\eta_{\alpha\nu}+h_{\alpha\nu})dx^{\alpha}dx^{\nu}  \]
yields to first order in $h_{\mu\nu}$
\begin{equation} \delta n^\mu |_i^f = -\eta^{\mu\nu}[h_{\nu 0} +
	h_{\nu j}n^j]_i^f  +{\eta^{\mu\sigma}\over 2}
   \int_i^fh_{\rho\nu,\sigma}n^{\rho}n^{\nu}d\lambda  \; ,
\label{2deltan} \end{equation}
where the integral is along the unperturbed photon trajectory from some
initial spacetime point $p_i$ to the endpoint $p_f$. On the right hand
side, unperturbed values for $n^\mu$ can be inserted.
Starting from this general relation, one obtains the Sachs Wolfe effect
by discussing the perturbation of $(n\cdot u)$, where $u$ is
the four velocity of an observer \cite{SW,d90,d93}.

The direction of a light ray with respect to an observer moving with four
velocity $u$ is given by the direction of the spacelike vector
\begin{equation} n_{(3)} =  (n + (u\cdot n)u)/|u\cdot n|  ~,
	  \label{2n3} \end{equation}
which lives on the sub--space of tangent space normal to $u$.
We now want to define the deflection of a light ray. At the point of
emission, the photon direction as measured by an observer moving according
to the velocity field $u$ is given by $n_{(3)}(p_i)$.
Correspondingly, at the point of detection it is given by
$n_{(3)}(p_f)$. We have to compare these two vectors
at different points. We do not want to switch off the perturbation,
 since this is not a gauge--invariant concept.
(There exist different averaging procedures, e.g. over different spacelike
hypersurfaces, which lead to slightly different Friedmann backgrounds.
Actually the difference in the obtained backgrounds can be of the same order
of magnitude as the amplitude of the perturbations  \cite{d93}).
We thus let the observer transport her frame of reference from the point of
emission to the point of detection and compare the direction of
$n_{(3)}(p_f)$ with respect to the transported frame.

We do not want to restrict the observer to move on a geodesic.
The correct way to transport a
 direction along a non geodesic curve is Fermi transport (see, e.g.
\cite{St}). The Fermi transport equation is given by
\begin{equation} \nabla_un^{||}_{(3)} =(n^{||}_{(3)}\nabla_uu)u
	~.\label{2Fer} \end{equation}
Here, we denote by $n_{(3)}^{||}$ the transported direction of emission
which coincides with $n_{(3)}$ at the point of emission.
This equation is uniquely specified by the requirement that Fermi transport
should conserve scalar products with $u$ and scalar products of
vectors normal to $u$. Therefore, angles and lengthes in the subspace of
tangent space normal to $u$ are conserved.

If an observer Fermi transports her frame of
reference with respect to which angles are measured, she
considers a light ray as not being deflected if
 $n^{||}_{(3)}(p_f)$ is parallel to $ n_{(3)}(p_f)$. The difference between
the direction of these two vectors is thus the deflection:
\begin{equation} \phi e = \left[ n_{(3)} - (n^{||}_{(3)}\cdot n_{(3)})
    n^{||}_{(3)}\right](p_f)
   ~. \label{2gendef}\end{equation}
Here $e$ is a spacelike unit vector normal to $u$ and normal to
$n^{||}_{(3)}$ which determines the direction of  deflection
and $\phi$ is the deflection angle. (Note that (\ref{2gendef}) is a
general formula for light deflection in an arbitrary gravitational field.
Up to this point we did not make any assumptions about the strength of
the field.) Clearly, due to this definition, the deflection angle will
in general depend on the path which the observer chooses to move from
$p_i$ to $p_f$. But we shall see in the following, that at least in cases
where the gravitational field originates from a spatially confined
mass distribution,
the observer  can always move on a path far away from all masses, so that
the path dependent contribution becomes negligible.

For a spherically symmetric problem, $e$ is uniquely determined by the
above orthogonality conditions since
the path of light rays is confined to the plane normal to the angular
momentum. In the general case, when the angular momentum of photons
is not conserved,
$e$ still has one degree of freedom. We now calculate $\phi e$
perturbatively.
Let us  define the perturbed quantities:
\begin{eqnarray*}
 n = (1, \mbox{\boldmath $n$}) + \delta n  ~~\mbox{ with }~~
	\mbox{\boldmath $n$}^2 = 1~; &~&
 u = (1,0) + \delta u ~=~ (1+{1\over 2}h_{00}, \mbox{\boldmath $v$}) \\
 n_{(3)} = (0,\mbox{\boldmath $n$}) + \delta n_{(3)}   ~~\mbox{ and}&&
 n^{||}_{(3)} = (0,\mbox{\boldmath $n$}) + \delta n^{||}_{(3)}
	~.\end{eqnarray*}
The perturbation $\delta n$ is given in (\ref{2deltan}). From (\ref{2n3}) we
obtain
\begin{equation} \delta n_{(3)} = \epsilon u +
     \epsilon(0,\mbox{\boldmath $n$})+ \delta n -\delta u ~,~~~
	\mbox{ with}~~~
 \epsilon = [n^iv^i -\delta n^0 +{1\over 2}h_{00}+n^ih_{i0}]
	~.  \label{2den3} \end{equation}
The Fermi transport equation (\ref{2Fer}) yields
\begin{eqnarray} \delta (n^{||}_{(3)})^0 &=& n^i(h_{i0}+v_i) \\
  \delta (n^{||}_{(3)})^j &=&  -{1\over 2}[h_{lj}n^l |_i^f+
    \int_i^fdt(h_{j0,l} - h_{l0,j})n^l] \label{2denp}~.\end{eqnarray}
Inserting (\ref{2den3}--\ref{2denp}) into (\ref{2gendef}) leads to
\begin{eqnarray}  \phi e^0 = 0~,  &&
  \phi e^i= \delta^i-(\mbox{\boldmath $\delta\cdot n$})n^i~,
	~~\mbox{ with} \label{2ei} \\
    \delta_j &=& [\delta n_j -v_j +{1\over 2} h_{jk}n^k]|_i^ f+{1\over 2}
   \int_i^f(h_{0j,k}-h_{0k,j})n^kdt ~. \label{2dej} \end{eqnarray}
This quantitiy is observable and thus gauge--invariant.
The last integral has to be performed along the path of the observer.
For a finite mass
distribution, at distance $R$ away from all masses, $h_{i0}<M/R$, therefore
 $h_{0i},_j <M/R^2$. On the other hand the length of the path is of order
$R$, so that we find
\[  \int_i^f(h_{0j,k}-h_{0k,j})n^kdt \le {M\over Rv} ~ .\]
The observer can thus always choose a path far away from the mass
distribution so that this  term can be neglected.

We now want to express the general formula (\ref{2ei},\ref{2dej}) in terms
of gauge--invariant variables for scalar, vector and tensor type
perturbations.
The most general form for {\bf scalar} perturbation of the metric is
given by
\begin{equation} \left(h_{\mu\nu}\right) = \left(\begin{array}{ll}
		2A & 2B,_i \\
                2B,_i & 2(H_L-{1\over 3}\triangle H)\delta_{ij}
                                +2H,_{ij} \end{array}\right)
 \label{scalar} \end{equation}
and a {\bf scalar} velocity field can be derived from a potential,
$v_i = -v,_i$.
One can show that the following combinations of these variables are
gauge--invariant \cite{Ba,KS,d90}
\begin{equation} V = v-\dot{H}~;~~ \Psi = A+\dot{H}-B~,~~
	\Phi = H_L-{(1/ 3)}H~. \end{equation}
The variables $\Phi$ and $\Psi$ are the so called Bardeen potentials.
(In  Newtonian approximation with Newtonian potential $\varphi$ one finds
$\Psi = -\Phi = \varphi ~. $)
 With ansatz (\ref{scalar}) equation (\ref{2dej}) leads to
\begin{equation}
(\delta_j)^{(S)} = V_{,j}|_i^f +\int_i^f(\Phi-\Psi)_{,j}d\lambda
 \label{2defs} ~.\end{equation}
For spherically symmetric perturbations, where $e$ is uniquely defined,
we can write this result in the form
\begin{equation} \phi^{(S)} = V_{,i}e^i|_i^f +
	\int_i^f(\Phi-\Psi)_{,i}e^id\lambda ~.
 \label{2ldefl} \end{equation}
The first term here denotes special relativistic spherical aberration.
 The second term represents  gravitational light deflection, which
since it is conformally invariant, only depends on the Weyl part of
the curvature. For scalar perturbations, the amplitude of the Weyl tensor
 is given by $(\Psi-\Phi)$ \cite{d93,BDE}.
As an  easy test we insert the Schwarzschild weak field approximation:
$\Psi = -\Phi = -{GM/ r}$. The unperturbed geodesic is a straight line,
$ x = (\lambda, \mbox{\boldmath $n$}\lambda + \mbox{\boldmath $e$}b)$,
where $b$ denotes the impact parameter
of the photon. Inserting this into (\ref{2ldefl}) yields
Einstein's well known result: $\phi^{(S)} = {4GM/ b}$.

The most general ansatz for {\bf vector} perturbations  is
\begin{equation} \left(h_{\mu\nu}\right) = \left(\begin{array}{ll} 0 &
	2B_i \\
                              2B_i & H_{i,j}+ H_{j,i}
                               \end{array}\right),~~~ v^i~,
\end{equation}
where $B_i$, $H_i$ and $v_i$ are divergence free vector fields.

The following combinations of these variables are gauge--invariant
\cite{KS,d93}
\begin{equation} \Omega_i = v_i - B_i~, ~~~~ \sigma_i =
	\dot{H}_i - B_i ~.\end{equation}
Inserting  these definitions in (\ref{2dej}), we obtain
\begin{equation} (\delta_j)^{(V)} = \Omega_{j}|_i^f -{1\over 2}[ \int_i^f
   (\sigma_{j,k}-\sigma_{k,j})n^kdt +\int_i^f\sigma_{k,j}n^kd\lambda]  ~.
\label{2defv} \end{equation}
This result can be expressed in three dimensional notation  as follows:
\begin{eqnarray*} \lefteqn{\phi\mbox{\boldmath $ e$}=
 \mbox{\boldmath $\delta$}-(\mbox{\boldmath $\delta\cdot n$})
	\mbox{\boldmath $n$} =}  \\
 && -(\mbox{\boldmath $\Omega\wedge n$})\wedge \mbox{\boldmath $n$}|_i^f
	+ {1\over 2}\int_i^f
  (\mbox{\boldmath $\nabla$}\wedge\mbox{\boldmath $\sigma$})\wedge
	\mbox{\boldmath $n$}dt - \int_i^f
 (\mbox{\boldmath $\nabla$}(\mbox{\boldmath $\sigma\cdot n$})\wedge
\mbox{\boldmath $n$})\wedge\mbox{\boldmath $n$}d\lambda ~. \end{eqnarray*}
The first term is again a special relativistic  "frame dragging" effect.
The second term is the change of frame due to the gravitational
field along the path of the observer.  The third term gives the
gravitational light deflection. (Special relativistic Thomas
precession is not recovered in linear perturbation theory
 since it is of order $v^2$.)

This formula can be used to obtain in first order the light deflection
in the vicinity of a rotating neutron star or a Kerr black hole:\\
In suitable coordinates the metric of a Kerr black hole with mass $M$ and
angular momentum $M\mbox{\boldmath $a$}$ can be
approximated by \cite{St}
\begin{eqnarray}   g_{00} = -(1-2m/r) + {\cal O}(r^{-3}) &&
  g_{0i} = 2\epsilon_{ijk}{s^jx^k/ r^3}+ {\cal O}(r^{-3}) \\
  g_{ij} &=& \delta_{ij}(1-{2m/ r}) + {\cal O}(r^{-3})  ~ ,\end{eqnarray}
with $m=GM$, $\mbox{\boldmath $s$}=GM\mbox{\boldmath $a$}$ and
	$|\mbox{\boldmath $a$}| < GM$
($|\mbox{\boldmath $a$}|=GM$ represents the extrem Kerr solution). Therefore
\begin{equation} \mbox{\boldmath $\sigma$} = -\mbox{\boldmath $B$}
  = -({2/ r^3})\mbox{\boldmath $s$}\wedge\mbox{\boldmath $r$}
	~. \end{equation}
We consider an observer which emits a light ray with impact vector
$b\mbox{\boldmath $e$}$ in direction {\boldmath $n$} at infinity
and detects it at infinity on the other side of the black hole. She moves
to the point of detection in a wide circle around the black hole, so that
the contribution from the path of the observer can be neglected.
Light deflection is then given by
\begin{equation}\mbox{\boldmath $\phi$}^V = {4GM\over b^2}
	[\mbox{\boldmath $ a$}
	\wedge\mbox{\boldmath $ n$} +
2((\mbox{\boldmath $a$}\wedge\mbox{\boldmath $n$})\cdot
	\mbox{\boldmath $e$})\mbox{\boldmath $e$}].
\end{equation}
The size of the deflection angle is thus of the order
$ |\phi^{(V)}| \sim (a/b) |\phi^{(S)}|$ and is limited by
\[|\phi^{(V)}| \le {12GMa \over b^2} ={3a \over b}|\phi^{(S)}|  ~.\]
Only for extrem Kerr and very close encounters (where linear perturbation
theory is no longer valid since $b\sim GM$) the vector contribution is of
the same order of magnitude as the scalar term. For usual circumstances,
$a \ll b$, only the fact that the direction differs from
{\boldmath $e$} might
open the possibility of actually observing the vector contribution.

{\bf Tensor} perturbations  of the  metric  are of the form
\begin{equation} \left(h_{\mu\nu}\right) = \left(\begin{array}{ll} 0 & 0 \\
                               0& H_{ij}
                               \end{array}\right),
\end{equation}
where $H_{ij}$ is a traceless divergence free tensor field.
 The deflection angle  then becomes
\begin{equation} (\phi e_j)^{(T)} = -H_{jk}n^k|_i^f +\int_i^f
  (H_{lk,j} + \dot{H}_{kl}n_j)n^ln^kd\lambda ~.\label{2deft} \end{equation}
 Only the gravitational effects remain. The first
 contribution comes from the difference of the metric before and
after  passage of the gravitational wave. Usually this term is
negligible. The second term accumulates along the path of the photon.

We evaluate  (\ref{2deft}) for light deflection due to a
gravitational wave pulse for which the difference of the gravitational
field before and after passage of the pulse is negligible:
\begin{equation} \phi e_j =\int_i^f(H_{lk},_j+\dot{H}_{lk}n_j)n^kn^ld\lambda
	~. \end{equation}
During the crossing time of the photon, we approximate the pulse by
 a plane wave,
\[ H_{jl} = \Re(\epsilon_{jl}\exp(i(\mbox{\boldmath $k\cdot x$}-\omega t))
	~~\mbox{ with } ~~~
	\epsilon_{jl}k^l=\epsilon^l_l=0 ~. \]
For a photon with unperturbed trajectory
$x = (\lambda,\mbox{\boldmath $x$}_o
	+\lambda\mbox{\boldmath $n$})$,
we find the deflection angle
\[ |\phi| = \left\{\begin{array}{lll}
  \epsilon_{lm}n^ln^m{|\mbox{\boldmath $k$}-\omega\mbox{\boldmath $n$}|
	\over \omega-\mbox{\boldmath $k\cdot n$}}
	\cos(\alpha+(\mbox{\boldmath $k\cdot n$}-\omega)t) &
  \mbox{ for }& \mbox{\boldmath $k$} \neq \omega\mbox{\boldmath $n$} \\
 0 & \mbox{ for }& \mbox{\boldmath $k$} = \omega\mbox{\boldmath $n$}~.
\end{array} \right. \]
Setting $\mbox{\boldmath $n$} = (p/\omega)\mbox{\boldmath $k$} +
q\mbox{\boldmath $n$}_\perp$ with $\mbox{\boldmath $n$}_\perp^2=1$
and $p^2+q^2 = 1$, we have $\epsilon_{lm}n^ln^m =q^2\epsilon_\perp$, where
 $\epsilon_\perp =\epsilon_{ij}n_\perp^in_\perp^j$.
Inserting this above, we obtain
\begin{equation} \phi = \epsilon_\perp\sqrt{2}(1+p)\sqrt{1-p}
	\cos(\alpha +\omega(p-1)t)
	~. \label{3gw} \end{equation}
 The amplitude of the gravitational wave determines  $\epsilon_\perp$
and $p$ is given by the intersection angle of the photon
with the gravitational wave as explained above.

This effect for a gravitational wave from two coalescing black holes would
be quite remarkable: Since for this (most prominent) event
$\epsilon_\perp$ can be as large as $\approx 0.1(R_s/r)$, light rays passing
 the black hole with impact parameter  $b$ would be
deflected by the amount:
\[ \phi\approx 2''(10^4R_S/b) ~. \]
Setting the  the  source at distance $d_{LS}$ from the coalescing
black holes and at distance $d_S$
from us, we observe a deflection angle
\[ \beta  = \phi d_{LS}/d_S ~. \]
The best source candidates would thus be quasars for which $ d_{LS}/d_S$
 is of order unity for all  coalescing black holes with, say $z\le 1$.
In the vicinity of the black holes ($r\le 10R_S$), linear perturbation
theory is of course not applicable.  But in the wide range $10^7R_S>b>10R_S$
for radio sources,
and  $10^4R_S>b>10R_S$ for optical sources, our calculation
is valid and the result might be detectable.

A thorough investigation of the possibility of detecting  gravitational
waves of coalescing black holes out to cosmological distances by this effect
may be worth while. This possibility is also discussed in \cite{Fakir}.

We  have defined light deflection in a gauge--invariant,
operational way. In general, the result depends on the path along which the
observer transports her frame of reference from the point of emission
to the point of detection.
Our equations are derived in perturbed Minkowski space, but
since Friedmann Lema\^\i tre universes are conformally flat, they also
apply for them (in conformal coordinates!).

More applications of light deflection can be found in a review paper
on cosmological perturbation theory \cite{d93}. There
the formulas derived in this letter are used to calculate light deflection
from topological defects.

\end{document}